\documentclass{amsart}
\def\t{\hbox}

\def\f{\frac}

\def\q{\quad}
\def\p{\varphi}

\def\k{\kappa}

\def\a{\alpha}

\def\d{\delta}
\def\be{\begin{equation}}
\def\ee{\end{equation}}
\def\bea{\begin{eqnarray}}
\def\eea{\end{eqnarray}}
\def\ba{\begin{array}}
\def\ea{\end{array}}

\def\pr{\prime}

\theoremstyle{definition}

\theoremstyle{remark}

\numberwithin{equation}{section}

\newcommand{\abs}[1]{\lvert#1\rvert}

\newfont{\Bb}{msbm8 scaled\magstep{1}}
\newcommand{\rc}{\mbox{\Bb R}}

\usepackage{graphics}

\begin{document}

\title
[Stable Identification of Potentials]
{Stable Identification of Piecewise-Constant Potentials from 
Fixed-Energy Phase Shifts
}

\author{S. Gutman}
\address{Department of Mathematics\\ University of Oklahoma\\ Norman,
 OK 73019, USA}
\email{sgutman@ou.edu}

\author{Alexander G. RAMM}
\address{ Department of Mathematics\\ Kansas State University\\
Manhattan, Kansas 66506-2602, USA}
\email{ramm@math.ksu.edu}


\begin{abstract} 
An identification of a spherically symmetric potential by its phase
shifts is an important physical problem. Recent theoretical results 
assure that such a potential is uniquely defined by a sufficiently
large subset of its phase shifts at any one fixed energy level. However,
two different potentials can produce almost identical phase
shifts. That is, the inverse problem of the identification of a
potential from its phase shifts at one energy level $k^2$ is ill-posed,
and the reconstruction is unstable. In this paper we introduce a
quantitative measure $D(k)$ of this instability. 
The diameters of minimizing sets
 $D(k)$ are used to study the change in the stability  with the change of $k$,
 and the influence of noise on the identification. They are also used
in the stopping criterion for the nonlinear minimization 
method IRRS (Iterative Random Reduced Search).
IRRS combines probabilistic global and deterministic local
search methods and it is used for the numerical recovery of 
the potential by the set
of its phase shifts. The results of the identification for noiseless as
well as noise corrupted data are presented.
 
\end{abstract}

\maketitle



\section{Introduction}
Identification of a material from the measurements 
of its bombardment by particles
has long been of interest in physics. The results of such an 
experiments
are used to determine the phase shifts. Identification of a
potential by 
its phase shifts is an important physical problem, see
\cite{apagyi,cs,n,r3,r7}.
It has recently been shown that a
 sufficiently large infinite subset of the set of single 
energy phase shifts uniquely defines
a bounded compactly supported potential , \cite{r5}. However, in practice, only
a finite (and relatively small) subset of the phase shifts can be
determined from experimental data. It has been shown 
\cite{ars,gut5,gutmanramm2,rsmi} that given
such a finite set of shifts at an energy level $k^2$ one can find
several quite different potentials having practically the same phase
shifts at all $l$ (angular momenta) although Ramm's uniqueness theorem
\cite{r5} guarantees that the phase shifts $\delta_l:=\delta (l,k)$, known
at a
fixed $k>0$ for all values of $l\in {\mathcal L},$ such that
$\sum_{l\in {\mathcal L}, l\neq 0}\frac 1 l=\infty,$
determine uniquely a compactly supported spherically symmetric potential 
$q(r),$ such that $q(r)=0$ when $r>a>0$ and $\int_0^a
r^2|q(r)|^2dr<\infty$. 
Thus, the inverse problem of the recovery of the potential by its
phase shifts at one fixed energy level can be 
severely ill-posed, \cite{r8}. In this
paper we describe  a quantitative measure $D(k)$ of such an
instability, and an algorithm for its computation. $D(k)$ is the diameter of the minimizing set
defined in Section 3. The algorithm provides a method for its numerical estimate. 
Several numerical examples are presented in Section 4.

Let $q(x),\, x\in \rc^3,$ be a real-valued potential with compact
support. Let $R>0$ be a number
 such that $q(x)=0$ for $\abs{x} > R$. We also
assume that $q\in L^2(B_R)\,,\ B_R=\{x:\abs{x}\leq R, x\in \rc^3\}$. 
Let $S^2$ be the unit sphere, and $\alpha \in S^2$. For
a given energy $k>0$ the scattering solution $\psi(x,\alpha)$ is defined
as the solution of

\begin{equation}
\Delta \psi+k^2\psi-q(x)\psi= 0\,,\quad x \in \rc^3 
\end{equation}
satisfying the following asymptotic condition at infinity:

\begin{equation}
\psi=\psi_0+v,\quad \psi_0:=e^{ik\alpha\cdot x}\,,\quad \alpha\in S^2\,,
\end{equation}

\begin{equation}
\lim_{r\rightarrow\infty}\int_{\abs{x}=r}\left| \frac{\partial v}{\partial
r}-ikv\right|^2 ds=0\,.
\end{equation}

It can be shown, that 

\begin{equation}
\psi(x,\alpha)=\psi_0+A(\alpha',\alpha,k)
\frac{e^{ikr}}{r}+o\left(\frac{1}{r}\right)\,,\;
\text{as}\ \ r\rightarrow\infty\,, \quad
\frac{x}{r}=\alpha '\,.
\end{equation}

The function $A(\alpha',\alpha,k)$ is called 
the scattering amplitude, $\alpha$ and $\alpha'$
are the directions of the incident and scattered waves, and $k^2$
is the energy, see  \cite{n}, \cite{r3}.

For spherically symmetric scatterers $q(x)=q(r)$ the scattering
amplitude satisfies $A(\alpha',\alpha,k)=A(\alpha'\cdot\alpha,k)$. 
The converse is established in
\cite{r6}. Following \cite{rs}, the scattering amplitude
for $q=q(r)$ can be written as

\begin{equation}
A(\alpha',\alpha,k)=\sum^\infty_{l=0}\sum^l_{m=-
l}A_l(k)Y_{lm}(\alpha')\overline{Y_{lm}(\alpha)}\,,
\end{equation}

where $Y_{lm}$ are the spherical harmonics, normalized in
$L^2(S^2),$ and the bar denotes the
complex conjugate.

The fixed-energy phase shifts $-\pi<\delta_l\leq\pi$ 
($\delta_l=\delta(l,k)$,  $k>0$ is fixed) are related to $A_l(k)$ 
(see e.g., \cite{rs}) by the formula:

\begin{equation}
A_l(k)=\frac{4\pi}{k}e^{i\delta_l}\sin(\delta_l)\,.
\end{equation}

\section{Phase Shifts for Piecewise-Constant Potentials}

In general, phase shifts for a spherically symmetric potentials can be
computed by a variety of methods, e.g. by a variable phase method
described in \cite{calogero}. The computation involves solving
a nonlinear ODE for each phase shift. However, if the potential is compactly
supported and piecewise-continuous, 
a much simpler method described in \cite{ars} can be used.
It is summarized below.

Consider a finite set of points $0=r_0< r_1 <r_2<\dots<r_N=R$
and a piecewise-constant potential
\be\label{pot}
q(r)=q_i,\t{ on } [r_{i-1},r_i) \t{ for } 
i=1,\dots, N, \t{ and } q=0\t{ for }r\ge R.
\end{equation}

Denote $\k_i^2:=k^2-q_i$, where $i=1,\dots, N,$
and $k$ is some fixed positive number.
Consider the following problem for the radial Schr\"odinger equation:
\be
\f{d^2\p_l}{dr^2}+\Biggl(k^2-\f{l(l+1)}{r^2}\Biggl)\p_l=q\p_l,
\q \lim_{r\to 0}(2l+1)!!r^{-l-1}\p_l(r)=1,
\end{equation}

which we rewrite as:

\be\label{sc}
\f{d^2\p_l}{dr^2}+\Biggl(\k_i^2-\f{l(l+1)}{r^2}\Biggl)\p_l=0
\end{equation}

on the interval $r_{i-1}\le r < r_i$.
On $[r_{i-1},r_i)$ one has the following general solution of (\ref{sc})

\be
\p_l(r)=A_ij_l(\k_ir)+B_in_l(\k_ir),
\end{equation}
where
\be
j_l(kr)=\sqrt{\frac{\pi kr}2}J_{l+1/2}(kr)\,,\;
n_l(kr)=\sqrt{\frac{\pi kr}2}N_{l+1/2}(kr)\,
\end{equation}
and $J_l\,,\ N_l$ are the Bessel and Neumann functions.

We assume below that $\k_i$ does not vanish for all $i$. If $\k_i=0$
for some $i$, then (\ref{sc}) has
the solution

\be
\p_l(r)=A_ir^{l+1}+B_ir^{-l}\,,
\end{equation}
and our approach is still valid with obvious changes.

From the regularity of $\p_l$ at zero one gets $B_1=0$. Denote
$x_i=B_i/A_i$, then $x_1=0$. 
We are looking for the continuously differentiable solution $\p_l$.
Thus, the following interface conditions hold:
\be\ba{lcc}
A_ij_l(\k_ir_i)+B_in_l(\k_ir_i)=A_{i+1}j_l(\k_{i+1}r_{i})+
B_{i+1}n_l(\k_{i+1}r_{i}),\\ \\
\f{\k_i}{\k_{i+1}}[A_ij_l^\pr(\k_ir_i)+B_in_l^\pr(\k_ir_i)]=
A_{i+1}j_l^\pr(\k_{i+1}r_{i})+B_{i+1}n^\pr_l(\k_{i+1}r_{i}).
\end{array}
\end{equation}

The Wronskian $W(j_l(r),n_l(r))=1$, thus
\be\ba{lcc}
A_{i+1}=n^\pr_l(\k_{i+1}r_{i})[A_ij_l(\k_ir_i)+B_in_l(\k_ir_i)]
-\f{\k_i}{\k_{i+1}}n_l(\k_{i+1}r_{i})[A_ij_l^\pr(\k_ir_i)+B_in_l^\pr(\k_ir_i)],
\\
\\
B_{i+1}=\f{\k_i}{\k_{i+1}}j_l(\k_{i+1}r_{i})[A_ij_l^\pr(\k_ir_i)+
B_in_l^\pr(\k_ir_i)]
-j^\pr_l(\k_{i+1}r_{i})[A_ij_l(\k_ir_i)+B_in_l(\k_ir_i)].
\end{array}
\end{equation}
Therefore
\be
\begin{pmatrix}
A_{i+1}\\ B_{i+1}
\end{pmatrix}
=\f{1}{\k_{i+1}}
\begin{pmatrix}\a^i_{11} & \a^i_{12}\cr
\a^i_{21} & \a^i_{22}
\end{pmatrix}
\begin{pmatrix}A_{i}\cr B_{i}
\end{pmatrix},
\end{equation}
where the entries of the matrix $\alpha^i$ can be written explicitly:
\be\ba{lcc}
\a^i_{11}=\k_{i+1}j_l(\k_ir_i)n^\pr_l(\k_{i+1}r_{i})-
\k_{i}j_l^\pr(\k_ir_i)n_l(\k_{i+1}r_{i}),
\\
\\
\a^i_{12}=\k_{i+1}n_l(\k_ir_i)n^\pr_l(\k_{i+1}r_{i})-
\k_{i}n_l^\pr(\k_ir_i)n_l(\k_{i+1}r_{i}),
\\
\\
\a^i_{21}=\k_{i}j^\pr_l(\k_ir_i)j_l(\k_{i+1}r_{i})-
\k_{i+1}j_l(\k_ir_i)j_l^\pr(\k_{i+1}r_{i}),
\\
\\
\a^i_{22}=\k_{i}n_l^\pr(\k_{i}r_i)j_l(\k_{i+1}r_{i})-
\k_{i+1}n_l(\k_ir_i)j_l^\pr(\k_{i+1}r_{i}).
\ea
\end{equation}

Thus
\be\label{xk}
x_{i+1}=\f{\a^i_{21}+\a^i_{22}x_i}{\a^i_{11}+\a^i_{12}x_i},\q
x_i:=\f{B_i}{A_i}
\end{equation}

The phase shift $\d(k,l)$ is defined by

\be
\p_l(r)\sim{|F(k,l)|\over k^{l+1}}
\sin(kr-\frac{\pi l}{2}+
\delta(k,l))\quad r\to\infty\enspace ,
\end{equation}

where $F(k,l)$ is the Jost function.
For $r>R$

\be\label{as}
\p_l(r)=A_{N+1}j_l(kr)+B_{N+1}n_l(kr).
\end{equation}

From (\ref{as}) and the asymptotics
$j_l(kr)\sim\sin(kr-l\pi/2),\q n_l(kr)\sim-\cos(kr-l\pi/2)$,
$r\to\infty$, one gets:

\be\label{dkl}
\tan\delta(k,l)=-\f{B_{N+1}}{A_{N+1}}=-x_{N+1}\,.
\end{equation}
Finally, the phase shifts of the potential $q(r)$
are calculated by the formula:
\be\label{dklf}
\delta(k,l)=-\arctan x_{N+1}.
\end{equation}

Let $q_0(r)$ be a spherically symmetric piecewise-constant potential.
Fix an energy level $k$ and a sufficiently large $N$. Let
$\{\tilde\delta(k,l)\}_{l=1}^N$ be the set of its phase shifts.
Let $q(r)$ be another such potential, and let
 $\{\delta(k,l)\}_{l=1}^N$ be the set of its phase shifts.

The best fit to data
function $\Phi(q,k)$ is defined by

\begin{equation}\label{phi}
\Phi(q,k)=\frac{\sum^N_{l=1}\abs{\delta(k,l)-
\tilde\delta(k,l)}^2}
{\sum^N_{l=1}\abs{\tilde\delta(k,l)}^2}\,,
\end{equation}

The phase shifts are known to decay rapidly with $l$, see \cite{rai}.
Thus, for sufficiently large $N$, the function $\Phi$ is practically the same as
the one which would use all the shifts in (\ref{phi}).
The inverse problem of the reconstruction of the potential from its
fixed-energy phase shifts is reduced to the minimization of the objective function $\Phi$
over an appropriate admissible set.
 A minimization algorithm for this nonlinear problem providing
a stability estimate for the identification of the original potential
$q_0$ is given in the next Section.

\section{  Global and Local Minimization Methods}

We seek the potentials $q(r)$ in the class of piecewise-constant, spherically 
symmetric real-valued functions. Let the admissible set be
\begin{equation}\label{adm}
A_{adm} \subset \{(r_1,r_2,\dots,r_M,q_1,q_2,\dots,q_M)\ : \ 0\leq r_i\leq R\,,\ 
q_{low}\leq q_m \leq q_{high}\}\,,
\end{equation}

where the bounds $q_{low}$ and $q_{high}$
for the potentials, as well as the bound $M$ on
the expected number of layers are assumed to be known.

A configuration $(r_1,r_2,\dots,r_M,q_1,q_2,\dots,q_M)$ corresponds to
the potential

\begin{equation}
q(r)=q_m\,,\quad \text{for}\quad r_{m-1}\leq r<r_m\,,\quad 1\leq m\leq M\,, 
\end{equation}
where $r_0=0$ and $q(r)=0$ for $r\geq r_M=R$.

Note, that the admissible configurations must also satisfy

\begin{equation}\label{admr}
r_1\leq r_2\leq r_3 \leq\dots\leq r_M\,.
\end{equation}

Given an initial configuration $Q_0\in A_{adm}\subset
\rc^{2M},$ a local minimization method finds a
local minimum near $Q_0$. On the other hand, global minimization methods
explore the entire admissible set in order to find a global minimum of
the
objective function. While the local minimization is usually
deterministic, the majority of the global methods are probabilistic.
As usual for inverse scattering problems,
the best fit to
data function $\Phi$ has many local minima and points of 
nondifferentiability, see \cite{gut5}.
In this situation a combination of global probabilistic and local
deterministic methods
proved to be successful.

 In \cite{gutmanramm} such an algorithm
(the Hybrid Stochastic-Deterministic Method) has been 
applied for the identification 
of small subsurface particles,
given a set of surface measurements. The HSD method could be
described as a variation of a genetic algorithm and a local search
with reduction. In \cite{gut3}  
two global search algorithms in combination with
a special local search method were applied to the identification of 
piecewise-constant scatterers by 
acoustic type measurements. The Rinnooy Kan and Timmer's 
Multilevel Single-Linkage Method in a combination with a special Local 
Minimization Method
has been applied to the identification of piecewise-constant
spherically symmetric potentials by their phase shifts in \cite{gut5}.
We have used the Reduced Random Search Method in \cite{gutmanramm2} to
find
different potentials with practically the same phase shifts.

In this paper we use the Modified Reduced Random Search Method.
The important modification consists
of the consideration of minimizing sets and their
diameters as quantitative measures of the stability of the minimization
algorithm.

In a pure {\bf Random Search} method a batch $H$ of $L$ trial points is generated in
$A_{adm}$ using a uniformly distributed random variable. Then a local
search is started from each of these $L$ points. A local minimum with
the smallest value of $\Phi$ is declared to be the global one. 

In our
case $A_{adm}$ is a box in $\rc^{2M}$. The uniform random variable is
called $2M$ times to produce a point in this box (after the appropriate rescaling
 in each dimension). Finally, the obtained values of $r_i$ are rearranged in 
the ascending order to satisfy (\ref{admr}).

Since the Random Search method is computationally extremely inefficient, it is modified to reduce the number of
local searches. In the {\bf Reduced Sample Random Search} method one
uses only a certain fixed fraction  $\gamma$ of the original batch of $L$
points to proceed with the local searches. Typically, $L=5000$ and $\gamma=0.01$.
This reduced sample $H_{red}$ of
$\gamma L$ points  is chosen to contain the points with the smallest
$\gamma L$ values of $\Phi$ among the original batch. The local searches
are started from the points in this reduced sample. This way only the points that seem to
be in a neighborhood of the global minimum are used for an expensive local
minimization, and the computational time is not wasted on less
promising candidates.

 Let $H_{min}$ be the 
$\gamma L$ points obtained as the result of the local minimizations ($\gamma L=50$ in our computations). Let
$S_{min}$ be the subset of $H_{min}$ containing points
 $\{p_i\}$ with the smallest 
$\nu\gamma L$ ($0<\nu<1$, we used $\nu=0.1$) values in $H_{min}$. We call $S_{min}$ the
minimizing set. The choice of $\nu$ determines a representative sample
of global minimizers. If all these minimizers are close to each other,
then  the objective function $\Phi$ is not
flat near the global minimum. That is, the method identifies the minimum consistently. 
 To define this consistency in quantitative terms, let
$\|.\|$ be a norm in the admissible set.

Let
\begin{equation}\label{diam}
D=diam(S_{min})=\max\{\|p_i-p_j\|/d_{av}\ :\ p_i,p_j\in S_{min}\}\,,
\end{equation}
where $d_{av}$ is the average norm of the elements in $H_{min}$.
The normalization by $d_{av}$ is introduced to provide comparable results for different potentials.

Large $D$ indicates
that the found minimizers $p_i$ of $\Phi$ are far apart. 
In terms of the Inverse Problem, it means that the found solution is not
stable.
The diameter $D$ is a measure of such instability. A detailed
description of an iterative version of this algorithm is given at the
end of this Section. First, we discuss local minimization methods.

In our minimization algorithm the  Reduced Sample Random Search
method is
coupled with
a deterministic Local Minimization Method.
Numerical experiments show that the
objective function $\Phi$ is relatively well behaved in this problem:
while it contains many local minima and, at some points, $\Phi$ is not
differentiable, standard minimization methods work well here. 
A Newton-type method for the minimization of $\Phi$ is described in
\cite{ars}. We have chosen to use a variation of Powell's minimization
method which does not require the computation of the derivatives of the
objective function. Such method needs a minimization routine for 
a one-dimensional minimization of $\Phi$, which we do using a Bisection or a
Golden Rule method. See \cite{gut3} or \cite{gut5} for a complete
description of our method. 

Now we can describe our Basic Local Minimization
Method in $\rc^{2M}$, which is a modification 
of Powell's minimization method \cite{bre}. It is assumed here that the
starting position (configuration) $Q_0\in A_{adm}$ is suppied by the
procedure LMM (see below), and
the entry to LMM is provided by the global minimization part (IRRS).

\subsection*{Basic Local Minimization Method}
\begin{enumerate}

\item  Choose the set of directions $u_i\,,\;i=1,2,\dots,2M$ to be the standard basis
 in $\rc^{2M}$
\[
u_i=(0,0,\dots,1,\dots,0)\,,
\]
where $1$ is in the i-th place.
\item  Save your starting configuration supplied by LMM as $Q_0$ .
\item  For each $i=1,\dots,2M$ move from $Q_0$ along the line defined by $u_i$ 
 and find the point of minimum $Q_i^t$. This defines $2M$ temporary
points of minima.
\item  Re-index the directions $u_i$, 
so that (for the new indices) $\Phi(Q_1^t)\leq \Phi(Q_2^t)
\leq,\dots,\Phi(Q_{2M}^t)\leq\Phi(Q_0)$.

\item  For $i=1,\dots,2M$ move from 
$Q_{i-1}$ along the direction $u_i$ and find the point
of minimum $Q_i$.
\item  Set $v=Q_{2M}-Q_0$.
\item  Move from $Q_0$  along the direction $v$ and find the minimum. Call it
 $Q_0$ again. It replaces $Q_0$ from step 2.
\item Repeat the above steps until a stopping criterion is satisfied.
\end{enumerate}

Note, that we use the temporary points of minima $Q_i^t$ only to rearrange the
initial directions $u_i$ in a different order. The stopping criterion is
the same as the one in \cite[Subroutine Powell]{numrec}.

Still another refinement of the local phase is necessary
to produce a successful minimization. The admissible set $A_{adm}$, see
(\ref{adm})-(\ref{admr}), belongs to a $2M$ 
dimensional minimization space $\rc^{2M}$.
The dimension $2M$ of this space
is chosen a priori to be larger than $2N$, where $N$
is the number of layers in the original potential. We have chosen
$M=6$ in our numerical experiments. 
However, since the sought potential may have fewer than 
$M$ layers, we found that conducting searches in lower-dimensional subspaces
of $\rc^{2M}$ is essential for the local minimization phase.
A variation of the following "reduction" procedure has also been found 
to be necessary in \cite{gutmanramm} for the search of small subsurface objects, 
and in \cite{gut3} for the 
identification of multilayered scatterers.

If two adjacent layers in a potential
have  values $v_{i-1}$ and $v_i$ and the objective function $\Phi$
is not changed much when both layers are 
assigned the same  value $v_i$ (or $v_{i-1}$),
 then these two layers can be replaced with just one layer occupying their
place. The change in $\Phi$ is controlled by the parameter $\epsilon_r$.
We used $\epsilon_r=0.1$. This value, found from numerical
experiments, seems to provide the most consistent identification.
The minimization problem becomes
constrained to a lower dimensional subspace of $\rc^{2M}$ and the local
minimization is done in this subspace.

\subsection*{Reduction Procedure}

Let $\epsilon_r$ be a positive number.

\begin{enumerate}

\item  Save your starting configuration
$Q_0=(r_1,r_2,\dots,r_M,v_1,v_2,\dots,v_M)\in A_{adm}$
 and the value $\Phi(Q_0)$. Let the $(M+1)$-st 
layer be $L_{M+1}=\{r_M\leq |x| \leq
R\}$ and $v_{M+1}=0$.

\item  Let $2\leq i\leq M+1$. Replace $v_{i-1}$ 
in the layer $L_{i-1}$ by $v_i$. This defines a new configuration $Q_i^d$,
where the layers $L_{i-1}$ and $L_i$ are replaced with one new layer. Here $d$ stands
for the downward adjustment. Compute
$\Phi(Q_i^d)$ and the difference
$c_i^d=|\Phi(Q_0)-\Phi(Q_i^d)|$. Repeat for each layer in the original
configuration $Q_0$.

\item  Let $1\leq i\leq M$. Replace $v_{i+1}$ 
in the layer $L_{i+1}$ by $v_i$.  This defines a new configuration $Q_i^u$,
where the layers $L_i$ and $L_{i+1}$ are replaced with one new layer. Here $u$ stands
for the upward adjustment. Compute
$\Phi(Q_i^u)$ and the difference
$c_i^u=|\Phi(Q_0)-\Phi(Q_i^u)|$. Repeat for each layer in the original
configuration $Q_0$.

\item  Find the smallest among the numbers $c_i^d$ and $c_i^u$.
If this number is less than $\epsilon_r\Phi(Q_0)$, then implement the adjustment that
produced this number. The resulting new configuration has one less layer
than the original configuration $Q_0$.

\item Repeat the above steps until no further reduction in the number of
layers is occurring.

\end{enumerate}

Note, that an application of the Reduction Procedure may or may not
result in the actual reduction of the number of layers.

Finally, the entire Local Minimization Method {\bf (LMM)} consists of the
following:

\subsection*{Local Minimization Method (LMM)}
\begin{enumerate}

\item  Let your starting configuration supplied by IRRS be
$Q_0=(r_1,r_2,\dots,r_M,v_1,v_2,\dots,v_M)\in A_{adm}$.

\item  Apply the Reduction Procedure to $Q_0$, 
and obtain a reduced configuration
$Q_0^r$ containing $M^r$ layers.

\item  Apply the Basic Minimization Method in $A_{adm}\bigcap \rc^{2M^r}$ 
with the starting configuration $Q_0^r$, and obtain a configuration $Q_1$.

\item  Apply the Reduction Procedure to 
$Q_1$, and obtain a final reduced configuration
$Q_1^r$.

\end{enumerate}

As we have already mentioned, LMM is used as the local phase of the
global minimization. The global part is described as follows: 

\subsection*{Iterative Reduced Random Search (IRRS)} (at the $j-$th 
iteration).

Fix $0<\gamma, \nu, \beta <1,\ \epsilon>0$ and $j_{max}$.

\begin{enumerate}
\item  Generate another batch $H^j$ of $L$ trial points (configurations) in $A_{adm}$ using a uniform
random distribution.  

\item  Reduce $H^j$ to the reduced sample $H^j_{red}$ of $\gamma L$ points by selecting 
the points in $H^j$ with the smallest $\gamma L$ values of $\Phi$.

\item  Apply the Local Minimization Method (LMM) starting it at each point in $H^j_{red}$,
 and obtain the set $H^j_{min}$ consisting of the $\gamma L$ minimizers.

\item Combine $H^j_{min}$ with $H^{j-1}_{min}$ obtained at the previous
iteration. Let $S^j_{min}$ be the set of 
points from $H^j_{min}\cup H^{j-1}_{min}$ with the smallest
$\nu\gamma L$ values of $\Phi$. (Use $H^1_{min}$ for $j=1$).

\item  Compute the diameter $D^j$ of $S^j_{min}$ by 
$D^j=\max\{\|p_i-p_k\|/d_{av}\ :\ p_i,p_k\in S_{min}\}\,,$
where $d_{av}$ is the average norm of the $\gamma L$ elements of $H^j_{min}\cup H^{j-1}_{min}$
with the smallest values of $\Phi$.

\item (Stopping criterion). 

Let $p\in S^j_{min}$ be the point with the smallest value
of $\Phi$ in $S^j_{min}$ (the global minimizer).

If $D^j\leq\epsilon$, then stop. The global minimum is $p$. The
minimization is
stable.

If $D^j>\epsilon$ and $D^j\leq\beta D^{j-1}$, then return to step 1,
and do another iteration. (Stop, if the maximum number of iterations $j_{max}$ is
exceeded).

If $D^j>\epsilon$ and $D^j>\beta D^{j-1}$, then  stop. $p$ is the global minimum.
The minimization is unstable. The diameter $D^j$ is the measure of the
instability of the minimization.

\end{enumerate}

We used $\beta=0.95$, $\epsilon=0.01$ and $j_{max}=6$. The choice of
these and other parameters  ($L=5000,\, \gamma=0.01,\ \nu=0.1\, \epsilon_r=0.1$) is
dictated by their meaning in the algorithm and the comparative performance
of the program at their different values. As usual, some adjustment of parameters, stopping criteria
etc. is needed to achieve an optimal performance of the algorithm.

\section{Numerical Results}
We studied the performance of the algorithm for 4 
different potentials $q_i(r), i=1,2,3,4$.
In each case the following values of the parameters have been used. The
radius $R$ of the support of each $q_i$ was chosen to be $R=3.0$.
The admissible set $A_{adm}$ (\ref{adm}) was defined with $M=8$. The
Reduced Random Search parameters:
$L=5000\,,\;\gamma=0.01\,,\;\nu=0.1\,,\; \epsilon=0.01\,,\;\beta=0.95\,,j_{max}=6$.
The value $\epsilon_r=0.1$ was used in
the Reduction Procedure (see Section 3) during the local minimization phase.
The initial configurations were generated using a random 
number generator with seeds determined by the system time.
The run time was between 30 minutes to 2 hours on a 333 MHz PC, 
depending on the wave number $k$.
The number $N$ of the shifts used in (\ref{phi}) for the formation of
the objective function $\Phi(q)$ was determined from the condition
$\delta(k,0)>10^{-7}\delta(k,l)$. So, it was different for different
potentials $q_i$ and different wave numbers $k$. 
 The upper and lower bounds for the potentials $q_{low}$ and $q_{high}$
used in the definition of the admissible set $A_{adm}$ were chosen to
reflect the presumed a priori information about the potentials. While
one may attempt to deduce the bounds from the set of given data, i.e.
the phase shifts, it turns out, that an extensive admissible set may
lead to a nonuniqueness in the sought potential. This issue will be
studied elsewhere.

Let $q_1(r)$ be the following potential
\[
q_1(r)=\begin{cases} 
4.0 & 0\leq r < 0.3\\
1.0 & 0.3 \leq r < 1.0\\
-2.0 & 1.0 \leq r < 1.9\\
3.5 & 1.9 \leq r < 2.2\\
1.0 & 2.2 \leq r < 2.4\\
0.0 &  r \geq 2.4
\end{cases}
\]

\begin{figure}[fig1]
\vspace{5pc}
\includegraphics*{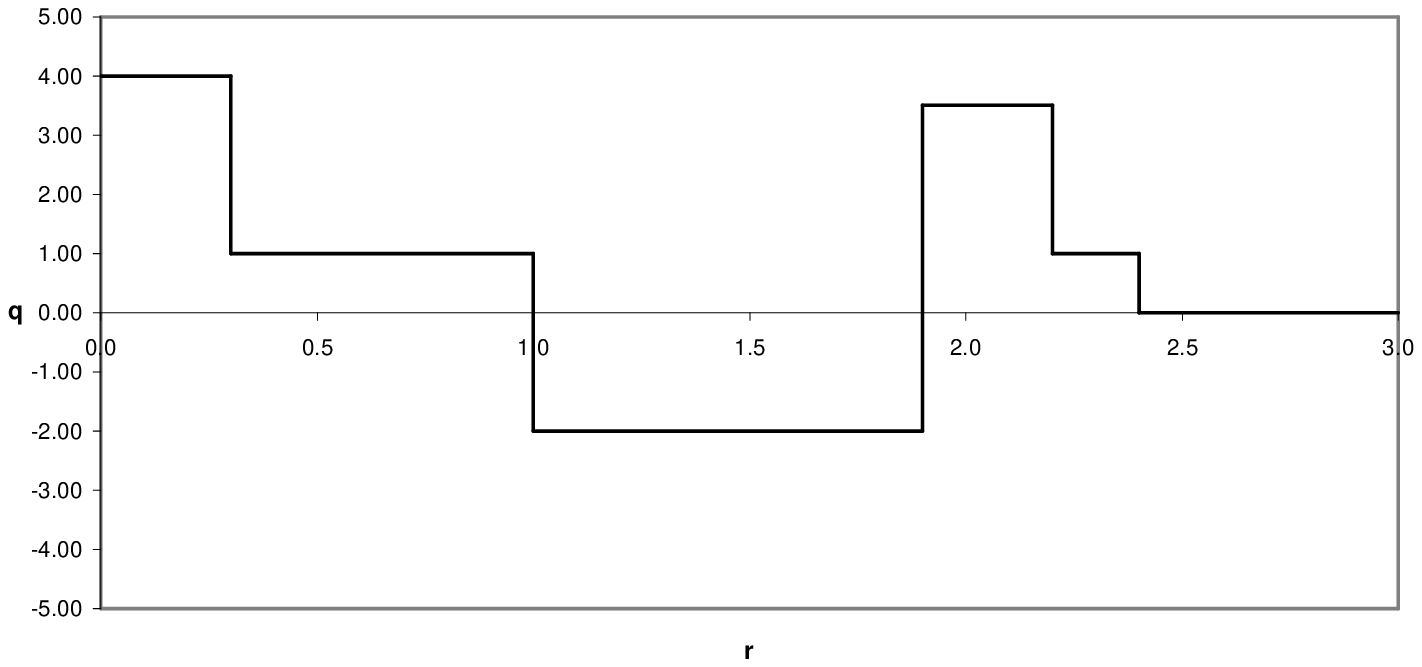}
\caption{Potential $q_1(r)$.}
\end{figure}

This potential is shown in Figure 1. The upper and lower bounds for the
 potential $q_{low}=-5.0$ and $q_{high}=5.0$ were used for all the wave
numbers $k=3,4,5,6,7,8,9$.


\begin{table}
\caption{Phase shifts of $q_1(r)$ for $k=9$.}

\begin{tabular}{r r| r r| r r}

\hline

$l$ & $\tilde\delta(k,l)$ & $l$ & $\tilde\delta(k,l)$ & $l$ &
$\tilde\delta(k,l)$ \\
\hline

 0 & -0.95151654D-01 & 12 & 0.43249567D-01 & 24 & -0.58868576D-03 \\
 1 & -0.59487863D-01 & 13 & 0.78575610D-01 & 25 &  -0.15074621D-03 \\ 
 2 & -0.30344479D-01 & 14 & 0.27082102D-01 & 26 &  -0.34641742D-04 \\
 3 & -0.36224576D-01 & 15 & -0.10281029D+00 & 27 &  -0.71940777D-05 \\
 4 &  0.14419664D-01 & 16 & -0.18261448D+00 & 28 & -0.13582152D-05 \\
 5 & -0.35167060D-01 & 17 & -0.17579851D+00 & 29 & -0.23433795D-06 \\
 6 &  0.38359584D-02 & 18 & -0.12758628D+00 & 30 & -0.37119174D-07 \\
 7 &  0.40280065D-01 & 19 & -0.76312741D-01 & 31 & -0.54203180D-08 \\
 8 &  0.45775379D-01 & 20 & -0.38650348D-01 & 32 & -0.73237693D-09\\
 9 &  0.99311592D-01 & 21 & -0.16752224D-01 & & \\
 10 &  0.93668476D-01 & 22 & -0.62688318D-02  &  &  \\
 11 &  0.32078999D-01 & 23 & -0.20460976D-02 &  &  \\

\hline
\end{tabular}

\end{table} 
 The phase shifts $\tilde\delta(k,l)$ for $k=9$ (computed as in
Section 2) are shown in Table 1.

The identification was attempted with 3 different noise levels $h$.
The levels are $h=0.0$ (no noise), $h=0.0001$ and $h=0.001$.
 More precisely, the noisy phase shifts $\delta_h(k,l)$ were obtained from
the exact phase shifts $\delta(k,l)$ by the formula

\[
\delta_h(k,l)=\delta(k,l)+(0.5-z)\cdot h\cdot \delta_{max}\,,
\]
where $\delta_{max}=max\{\abs{\delta_h(k,l)}\ :\ l=0,1,\dots,N\}$, and 
$z$ is the uniformly distributed on $[0,1]$ random variable.

The distance $d(p^1(r),p^2(r))$ for any two potentials in step 5 of the IRRS algorithm was
computed as 

\[
 d(p^1(r),p^2(r))=\|p^1(r)-p^2(r)\|\,
\]
where the norm is the $L_2$-norm in $\rc^3$.

The results of the identification algorithm (the diameters of the minimizing sets as the function of the
wave number $k$)
for the potential $q_1(r)$ are shown in Table 2 as well as in Figure 2.


\begin{table}
\caption{Diameters $D$ of minimizing sets for $q_1(r)$ at different noise levels $h$.}

\begin{tabular}{r r r r}

\hline

$k$ & $h=0.000$ & $h=0.0001$ & $h=0.001$ \\
\hline

3 &	0.886140 &	1.062616 &	0.870692\\
4 &	0.653900 &	0.629565 &	1.017525\\
5 &	0.636675 &	0.661651 &	0.754354\\
6 &	0.456157 &	0.326852 &	0.585782\\
7 &	0.008116 &	0.011184 &	0.066413\\
8 &	0.014142 &	0.004978 &	0.010448\\
9 &	0.007881 &	0.011849 &	0.022112 \\

\hline
\end{tabular}
\end{table} 

The diameter $D\leq 0.01$ indicates that the potentials in the minimizing
set are, practically, undistinguishable. That is, the identification is
stable for $k\geq 7$ and a low noise level.

\begin{figure}[fig2]
\vspace{5pc}
\includegraphics*{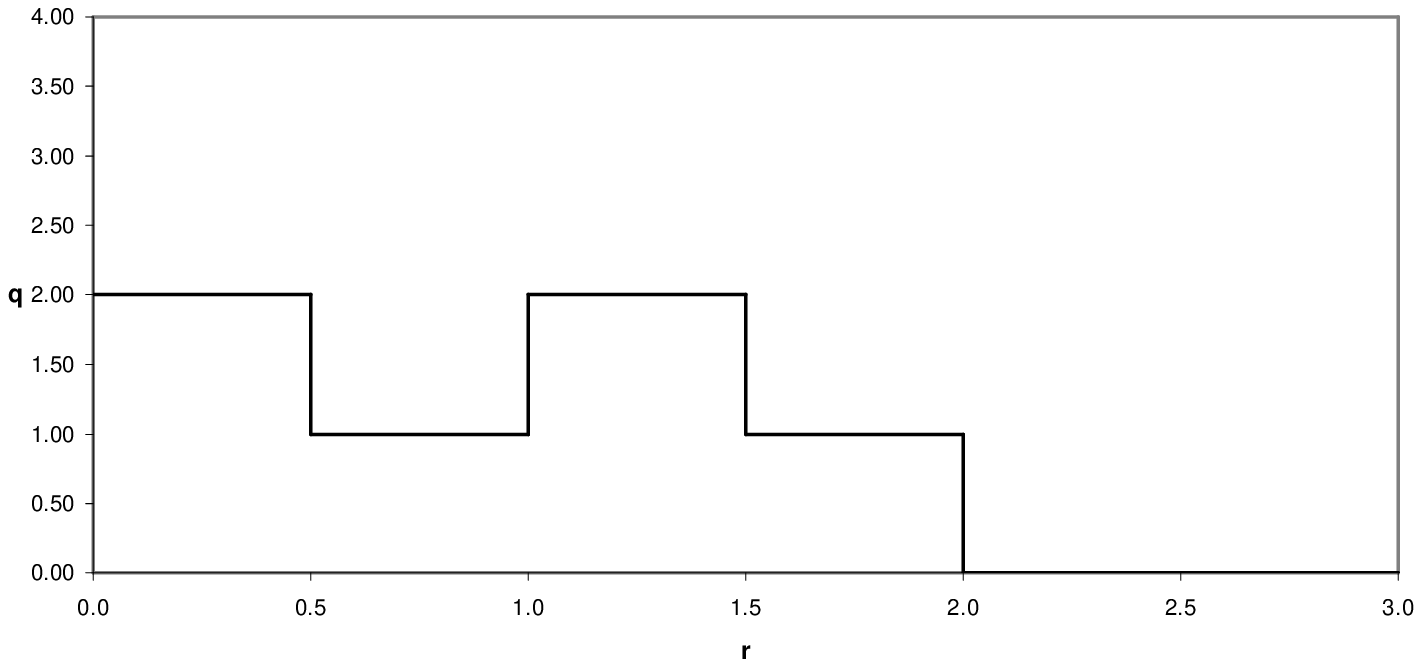}
\caption{Potential $q_2(r)$.}
\end{figure}

The second potential $q_2(r)$ is defined by
\[
q_2(r)=\begin{cases} 
2.0 & 0   \leq r < 0.5\\
1.0 & 0.5 \leq r < 1.0\\
2.0 & 1.0 \leq r < 1.5\\
1.0 & 1.5 \leq r < 2.0\\
0.0 &  r \geq 2.0
\end{cases}
\]

This positive 4-layer potential is shown in Figure 3. The upper and lower bounds for the
admissible set were the same as for $q_1$. The results of the
identification are shown in Table 3 and in Figure 4. The identification
is stable for $k\geq 6$ for low noise levels. It is becoming stable for
the noise level $h=0.001$ at $k=8$.


\begin{table}
\caption{Diameters $D$ of minimizing sets for $q_2(r)$ at different noise levels $h$.}

\begin{tabular}{r r r r}

\hline

$k$ & $h=0.000$ & $h=0.0001$ & $h=0.001$ \\
\hline

3 &	1.184754 &	1.865304 &	1.094802\\
4 &	1.435954 &	0.843646 &	0.887367\\
5 &	0.700282 &	0.684679 &	1.032036\\
6 &	0.000568 &	0.000990 &	0.693338\\
7 &	0.000373 &	0.001486 &	0.420729\\
8 &	0.000294 &	0.001435 &	0.003349\\
9 &	0.000190 &	0.001218 &	0.005308\\

\hline
\end{tabular}
\end{table}

\begin{figure}[fig3]
\vspace{5pc}
\includegraphics*{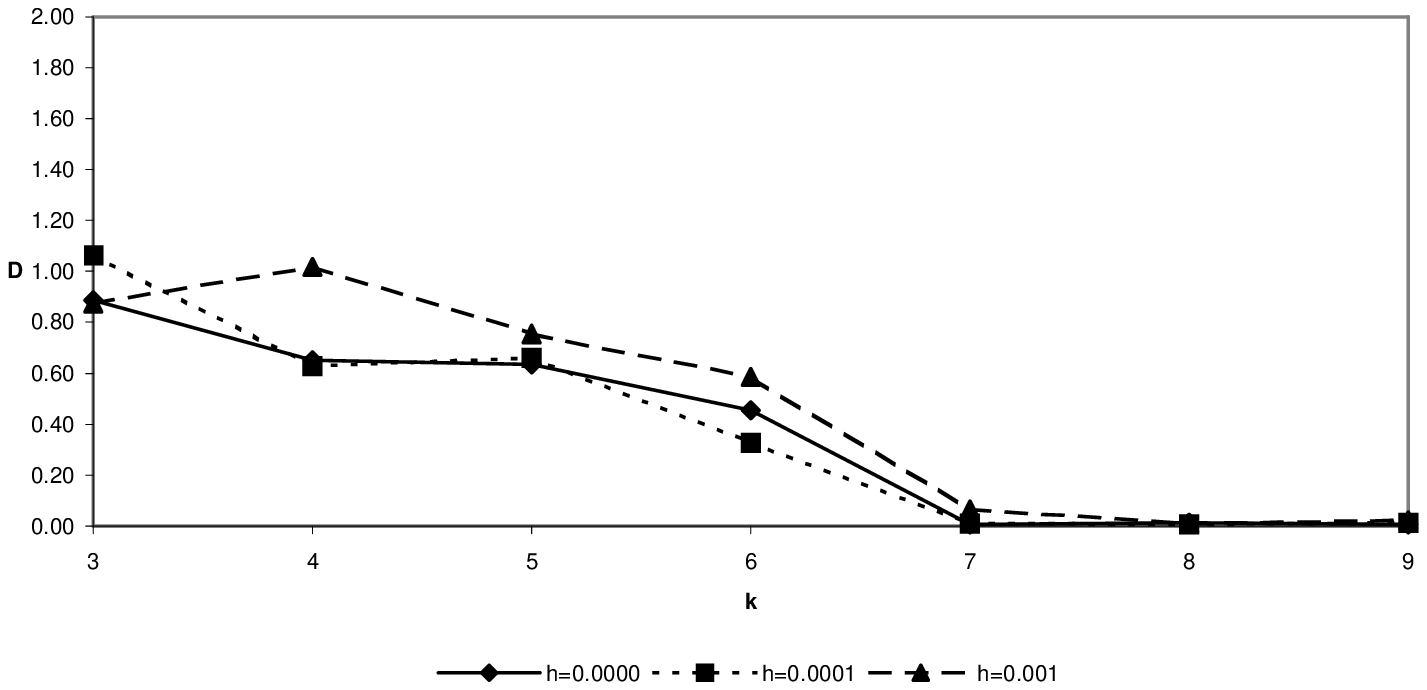}
\caption{Diameters $D$ of minimizing sets for $q_1(r)$ at different noise levels $h$.}
\end{figure}

\begin{figure}[fig4]
\vspace{5pc}
\includegraphics*{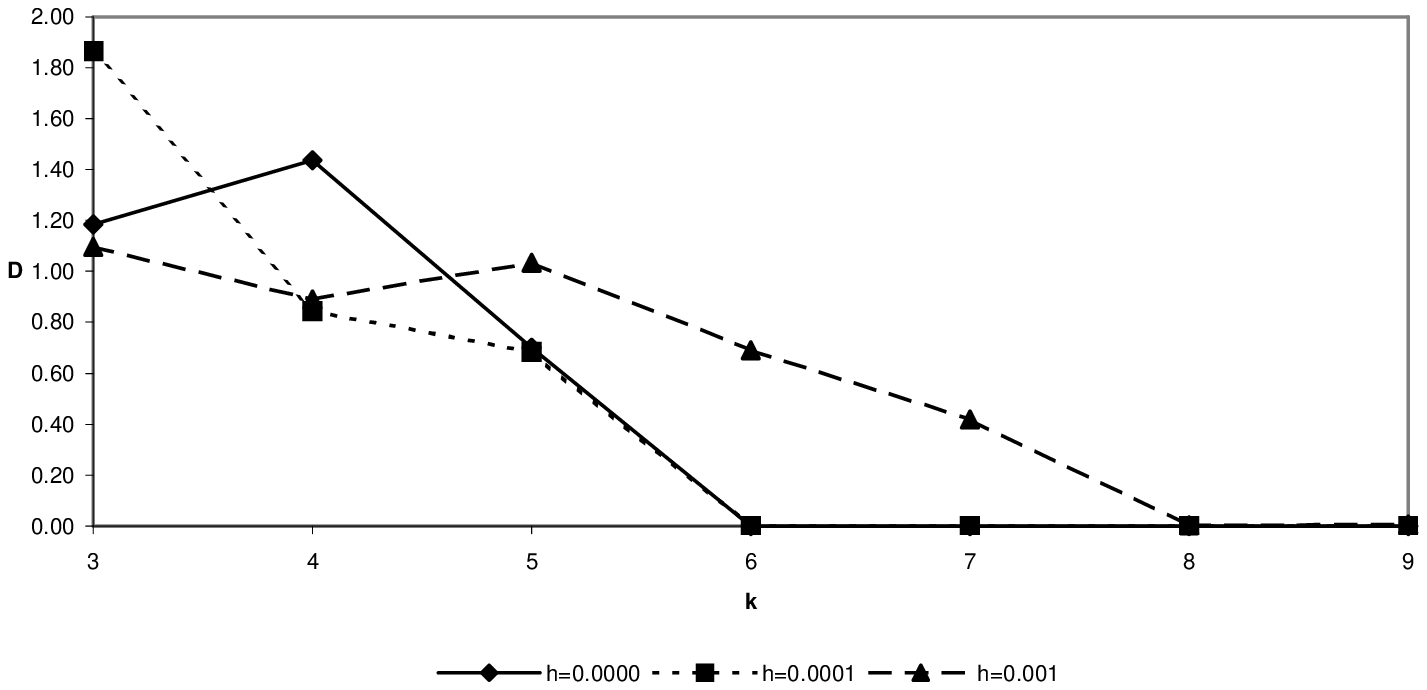}
\caption{Diameters $D$ of minimizing sets for $q_2(r)$ at different noise levels $h$.}
\end{figure}

We have also considered the identification for small potentials. In this
case the potential has to be identified from relatively small phase shifts.
We defined $q_3(r)=0.1q_2(r)$, and  $q_4(r)=0.01q_2(r)$. It turns out,
that the identification is comparable in quality to the ones above,
provided that reasonable a priori bounds for the potentials are supplied.
We used $q_{low}=-0.5$ and $q_{high}=0.5$ for the identification of potential
$q_3(r)$, and $q_{low}=-0.05$ and $q_{high}=0.05$ for potential
$q_4(r)$. The results of the identification are shown in Tables 4 and 5, as well as 
in Figures 5 and 6.


\begin{table}
\caption{Diameters $D$ of minimizing sets for $q_3=0.1q_2$ at different noise levels $h$.}

\begin{tabular}{r r r r}

\hline

$k$ & $h=0.000$ & $h=0.0001$ & $h=0.001$ \\
\hline

3 &	0.914191 &	0.958988 &	1.271533\\
4 &	0.649307 &	1.432330 &	0.571411\\
5 &	0.258550 &	0.425632 &	0.599525\\
6 &	0.000541 &	0.001361 &	0.318754\\
7 &	0.000373 &	0.211909 &	0.004215\\
8 &	0.000295 &	0.002696 &	0.006563\\
9 &	0.000170 &	0.003122 &	0.007665\\

\hline
\end{tabular}
\end{table}


\begin{table}
\caption{Diameters $D$ of minimizing sets for $q_4=0.01q_2$ at different noise levels $h$.}

\begin{tabular}{r r r r}

\hline

$k$ & $h=0.000$ & $h=0.0001$ & $h=0.001$ \\
\hline

3 &	1.233750 &	0.882006 &	1.681609\\
4 &	1.249050 &	0.891463 &	0.683901\\
5 &	0.676837 &	1.082434 &	0.309618\\
6 &	0.000565 &	0.001337 &	0.136704\\
7 &	0.000270 &	0.001288 &	0.004241\\
8 &	0.000523 &	0.001006 &	0.002867\\
9 &	0.000358 &	0.000991 &	0.005288\\

\hline
\end{tabular}
\end{table}

\begin{figure}[fig5]
\vspace{5pc}
\includegraphics*{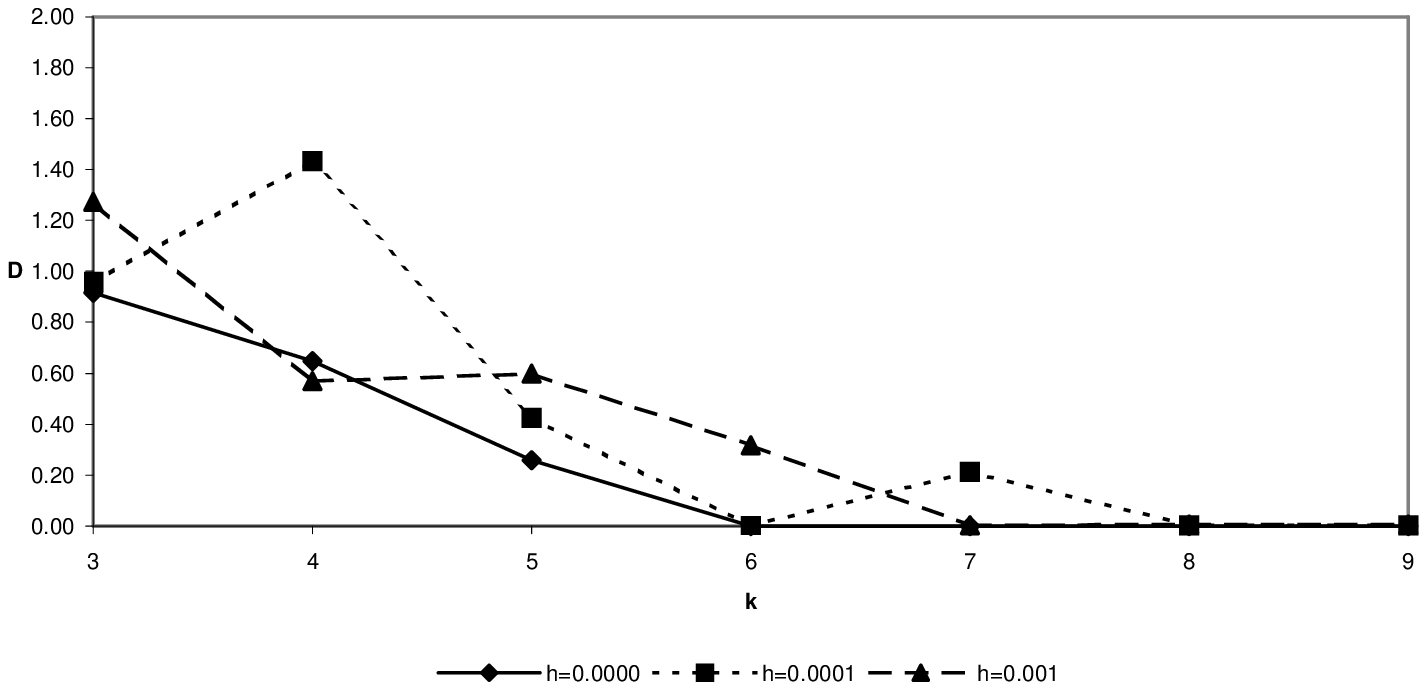}
\caption{Diameters $D$ of minimizing sets for $q_3=0.1q_2$ at different noise levels $h$.}
\end{figure}

\begin{figure}[fig6]
\vspace{5pc}
\includegraphics*{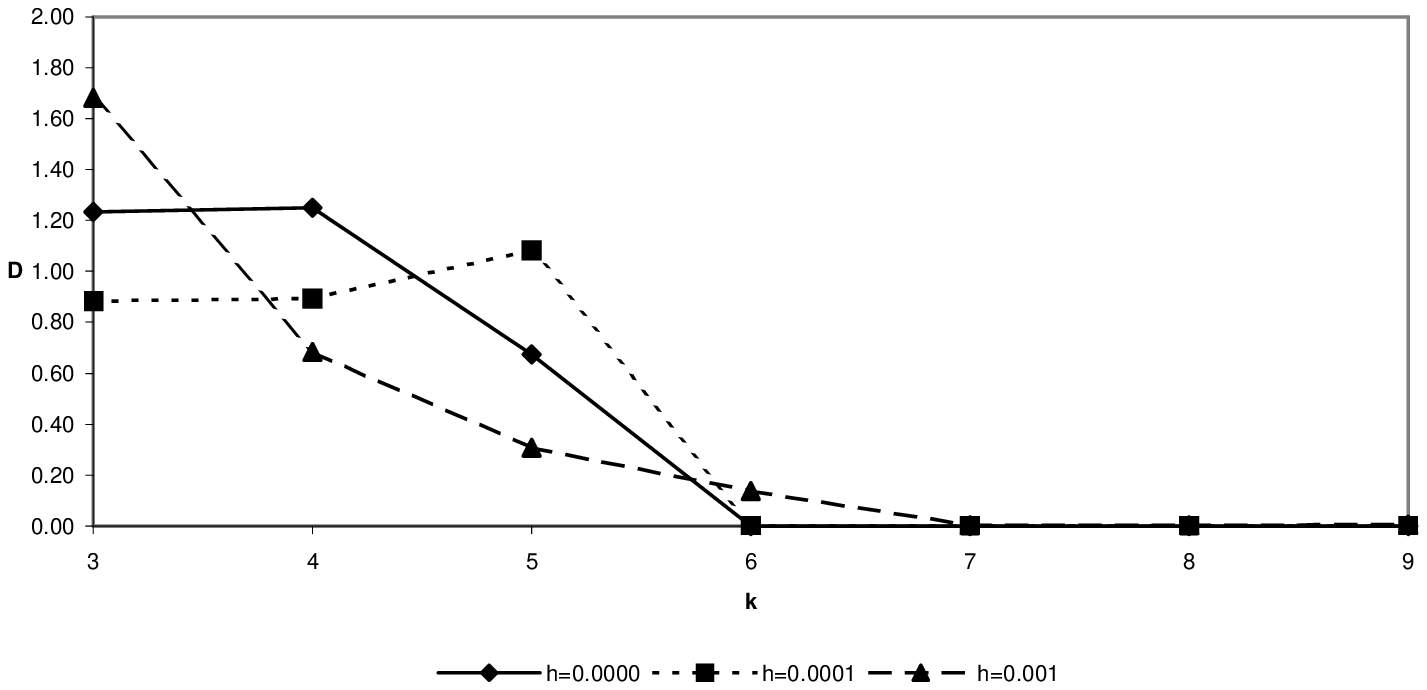}
\caption{Diameters $D$ of minimizing sets for $q_4=0.01q_2$ at different noise levels $h$.}
\end{figure}

\section{Conclusions}
Recovery of a spherically symmetric potential from its
fixed-energy phase shifts is a classical physical problem. 
Recent theoretical results \cite{r5}
assure that such a potential is uniquely defined by a sufficiently
large subset of its phase shifts at any one fixed energy level. However,
two different potentials can produce almost identical fixed-energy phase
shifts (\cite{ars}, \cite{gutmanramm2}). 
That is, the inverse problem of the identification of
the
potential by its fixed-energy phase shifts is very unstable. In this
paper we investigate the instability of the inversion by an Iterative
Reduced Random Search Method (IRRS). The diameter of the minimizing set
$D$ is introduced to provide a quantitative measure for the instability.
It also serves as the stopping criterion in the IRRS algorithm. The
results show, that for several types of piecewise-constant spherically
symmetric potentials the identification is becoming stable ($D\leq 0.01)$
for phase shifts measured at a higher energy level. It is also shown
that the introduction of a low noise level into the data does not
significantly degrades the identification. This method can serve as a
tool for experimentalists to determine if a particular set of phase
shifts would produce a stable identification of the underlying
potential, or a higher energy level should be used.

\end{document}